# Atypical electrical behavior of few layered WS$_2$ nanosheets based platform subject to heavy metal ion treatment

A Neog, S Deb, R Biswas*

Applied Optics and Photonics Lab, Department of Physics, Tezpur University, Tezpur-784028

*Corresponding author:rajib@tezu.ernet.in

**Abstract**

An atypical electrical behavior of WS$_2$ nanosheets deposited on Cu electrodes is reported here. The characteristic Raman peaks at 355cm$^{-1}$ and 421.8cm$^{-1}$ confirm the few-layer structure of WS$_2$. The addition of heavy metal ions of ~30µl on this platform results in non-ohmic behavior in I-V characteristics, accompanied by a dramatic rise of current from *n*A to *µ*A. Additionally, this atypical behavior is found to be reversible. Subsequent to removal of these ions from the nanosheets, it again exhibits normal ohmic I-V characteristics. It is envisioned that this unusual characteristic will pave way for more research in the sensing direction as well as relevant fields.

**Introduction**

The advent of semimetal Graphene [1] and its exotic behavior [2], [3] have inspired the exploration of other 2D materials. Transition Metal Dichalcogenides (TMDC) are some of the post-graphene 2D layered materials having identical structural properties with Graphene [4]. TMDCs are characterized by formula MX$_2$ where *M* represents transition metals including Molybdenum (Mo), Tungsten (W) etc. with X denoting chalcogenides. In monolayer, the transition metals are sandwiched between the chalcogens thereby possessing strong intra-layer covalent bond in between them. Inter-layer weak Van der Waals attraction facilitates the exfoliation of these materials into single and few layers, analogous to Graphene [5-6]. WS$_2$ and MoS$_2$ are some of the most studied semiconductors, belonging to the family of bulk TMDCs [6]**.** WS$_2$ outsmarts MoS$_2$ in terms of thermal stability [7]. The exfoliation of these materials into mono- or few layers considerably preserves the bulk properties due to quantum confinement [8]. Consequently, $WS_2$ nanosheets have drawn more attention for exceptional electronic properties, with wide use in optoelectronics and sensing applications [9-10]. Because of large scale synthesis along with higher solubility, they are used in biomedical applications [10]. Additionally, $WS_2$ nanosheets find extensive use in the fields of Catalyst[11], field-effect transistor[12], lithium ion battery[13], gas sensors[14],[15], thermal battery [16], super capacitors[17], humidity sensor[18], photosensor[19]. Recently, Zuo *et al* reported detection of heavy metal ions $Hg^{2+}$ and $Ag^+$ with the detection limit of 3.3 nM and 1.2 nM, respectively [20]. Again, Jia Ge *et al* have used $WS_2$ nanosheets for the detection of $Hg^{2+}$ ions with detection limit ~ 0.1 nM [21].

Generally, $WS_2$ possesses intrinsic S atoms being suitable coordination sites for certain heavy metal ions [22]. Very few studies address the basic affinity of $WS_2$ towards heavy metal ion treatment. Moreover, the I-V characteristics modulation of $WS_2$ due to heavy metal ion treatment is another unexplored domain. In the quest to look at these issues, we attempt to explore the sensing capability of $WS_2$ nanosheets towards the heavy metal ions through I-V characteristics with skipping other material functionalities. Accordingly, $WS_2$ nanosheets were uniformly drop casted on top of finger like structured copper electrodes.

**Experimental details**

*Chemicals and apparatus*

$WS_2$ (99.9%) was purchased from Sigma Aldrich. N-methly pyrrolidone, Acetone (99%), $CoCl_2$, $FeCl_3$ (98%) are purchased from Merck (USA). Deionised (DI) water was taken from a Millipore purification system. All reagents are of analytical grade and used as received without further purification. Ultrasonic bath Sonicator (UD100SH-2.8LQ), with power 100 W was used for sonication. Polypropylene centrifuge tubes (50ml) with conical bottom were used for sonication. For centrifugation, Remi centrifuge was used. Raman Spectra of $WS_2$ nanosheets were obtained using a Raman Spectrometer (RENISHAW). I-V characteristics were done through I-V characterizer (Ketheley 2400).

*Synthesis of $WS_2$ nanosheets*

Bulk $WS_2$ at a concentration of 1.5mg ml$^{-1}$ was dispersed in N-methyl pyrrolidone. The dispersed $WS_2$ has been placed in a medium power bath sonicator and sonicated for 3 hours with periodic shaking. Afterwards, the solutions were centrifuged for 15 minutes with 3000 rpm with eventual settling of $WS_2$ at the bottom and walls of the polypropylene tubes. The obtained pellets were again dispersed in DI water and centrifuged, followed by 12 hours of drying to attain the desired nanosheets.

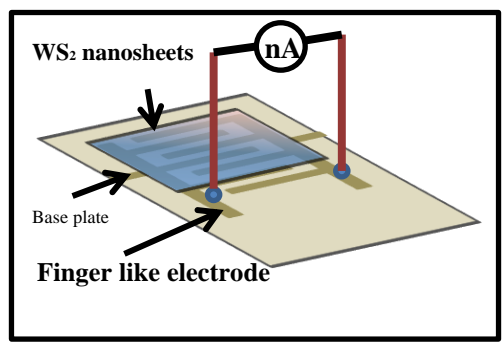 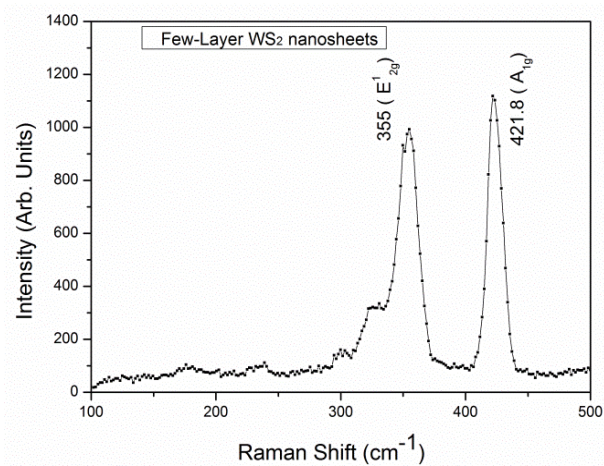

(a) (b)

Figure 1(a) Schematic of the fabricated finger like structured electrode device (b) Raman Spectra of atomically thin few layer $WS_2$ nanosheets

The as-synthesized $WS_2$ nanosheets were dispersed in DI water and ultrasonicated for 20 minutes and then drop casted on top of finger like Cu electrodes. The whole prototype was then kept in a clean chamber [see Figure 1(a)]. To investigate the effects of heavy metal ions on the drop casted $WS_2$ sheets, solution of ion was prepared. Concentrations of ions were maintained as

1ppm throughout the experiment. All I-V characterizations of the devices were done at room temperature.

**Result and Discussions**

*Raman Spectroscopy:*

Figure1 (b) shows the typical Raman Spectra of few layer $WS_2$ nanosheets, revealing two prominent peaks at 355cm$^{-1}$ and 421.8cm$^{-1}$ which correspond to $E^1_{2g}(\Gamma)$ and $A_{1g}(\Gamma)$ modes, respectively. It is also evident from the Raman Spectra that the intensity of $A_{1g}$ is higher than that of $E^1_{2g}$ mode with a Raman Shift of 66.8cm$^{-1}$. This is a spectral finger print of differentiating single layer(65cm$^{-1}$) from bulk $WS_2$ (68.6cm$^{-1}$)[14, 23], thus confirming the as-synthesized nanosheets to be of few layers [24]

*I-V analysis:*

As per previous report, the I-V characteristics of $WS_2$ film deposited on quartz [14] and $WS_2$ nanosheets drop casted on two electrode devices [19] show linear features with currents in ranges of *µ*A and *n*A, respectively. Our results corroborate those findings, exhibiting I-V characteristics of untreated $WS_2$ nanosheets based device to be linear in the voltage range of -9 to 9V (See Figure2). It is well evident that the scheme shows ohmic behavior. The current attained with respect to voltage variation emerges to be of *n*A and the maximum current attained was ~50nA.

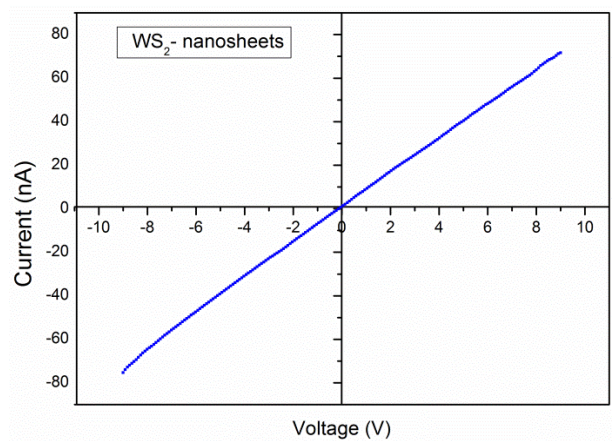

Figure 2. I-V characteristics of untreated $WS_2$ nanosheets

*Effect of heavy metal ions on I-V characteristics of WS₂ nanosheet based device:*

$Fe^{3+}$ and $Co^{2+}$ ions of concentration 1ppm were added to the WS$_2$ nanosheets to observe the impact of the ions on the films with the help of I-V characteristics as shown in Figures 3 a and b. The effects were monitored up to 20 minutes at an interval of 5 min. It can be clearly observed that, as soon as the heavy metal ions were impinged on WS$_2$, the current rises from *n*A to *μ*A with an abrupt change in the electrical behavior. In other words, response is more or less found to be atypical which is illustrated in Figure 3 a and b corresponding to treatment with Fe$^{3+}$ and Co$^{2+}$ ions. Reversibly, removal of ions from this WS$_2$ nanosheet restores the original characteristics as depicted in Figure 3 (c). The maximum current attained after removal was ~71-75$n$A. Though the current obtained in the latter is less, but the change is negligible. This reversible property of the device signifies the possibility of their use in different electrochemical sensing applications.

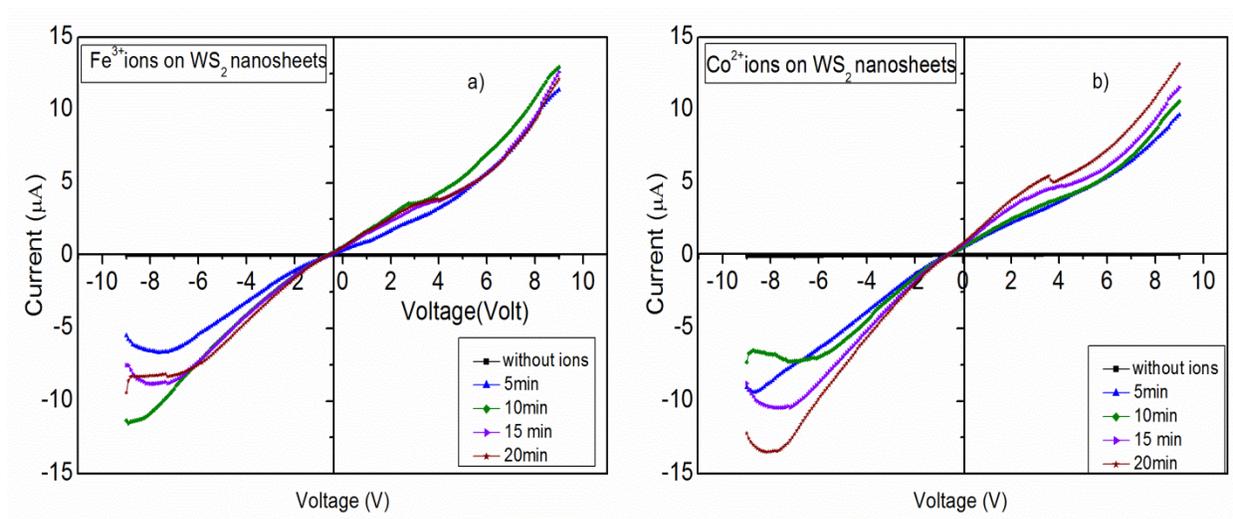

(a)                                (b)

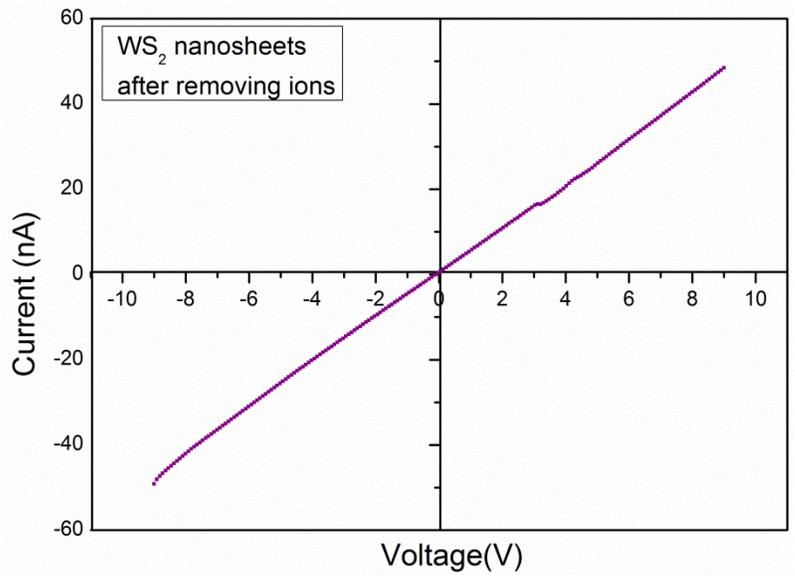

(c)

Figure 3. I-V characteristics of WS$_2$ nanosheets after adding a) Fe$^{3+}$ ions b) Co$^{2+}$ ions c) after removal of ions

**Conclusion:**

In summary, we report atypical electrical behavior of WS$_2$ nanosheet on Cu platform. The cost-effective scheme shows ohmic behavior in absence of heavy metal ion. From IV characteristics, the current attained in pristine conditions is found to be $n$A range. However, subject to concentration of heavy metal ion as low as ~30µl, the I-V characteristics shows atypical properties (non-ohmic). Subsequent removal of ions from the scheme restores original characteristics. This reversibility poses to be a unique platform for fabrication of sensitive devices based on I-V characteristics. Further addition of functionalities to this device can make it act as a selective heavy metal ion sensor.